\documentclass{aa}
\usepackage{graphics,times}

\begin{document}

   \title{From the Circumnuclear Disk in the Galactic Center to thick,
          obscuring tori of AGNs}

   \titlerunning{From the CND in the GC to obscuring tori in AGNs}

   \authorrunning{Vollmer et al.}

   \author{B.~Vollmer\inst{1,2}, T.~Beckert\inst{2}, \and
           W.J.~Duschl\inst{3,2}}

   \offprints{T.~Beckert, e-mail: tbeckert@mpifr-bonn.mpg.de}

   \institute{CDS, Observatoire astronomique de Strasbourg, 
       UMR 7550, 11 rue de 
       l'Universit\'e, 67000 Strasbourg, France \and
	      Max-Planck-Institut f\"ur Radioastronomie,
              Auf dem H\"ugel 69, 53121 Bonn, Germany \and
              Institut f\"ur Theoretische Astrophysik
              der Universit\"at Heidelberg,
              Tiergartenstra{\ss}e 15, 69121 Heidelberg, Germany}

   \date{Received / Accepted}

\abstract{
We compare three different models of clumpy gas disk and show that
the Circumnuclear Disk (CND) in the Galactic Center and a putative,
geometrically thick, obscuring torus are best explained by a collisional
model consisting of quasi-stable, self-gravitating clouds. Kinetic energy
of clouds is gained by mass inflow and dissipated in cloud collisions.
The collisions give rise to a viscosity in a spatially averaged
gas dynamical picture, which connects them to angular momentum transport
and mass inflow. It is found that CND and torus share the same gas physics
in our description, where the mass of clouds is  20 - 50 $M_\odot$ and
their density is close to the limit of disruption by tidal shear. 
We show that the difference between a transparent CND and an obscuring torus 
is the gas mass and the velocity dispersion of the clouds.  A change in gas
supply and the dissipation of kinetic energy can turn a torus into a
CND-like structure and vice versa.
Any massive torus will naturally lead to sufficiently high mass accretion
rates to feed a luminous AGN. For a geometrically thick torus to
obscure the view to the center even super-Eddington accretions rates
with respect to the central black hole are required.

 }

\maketitle

\section{Introduction \label{sec:intro}}

The central black hole of the Galaxy is surrounded by several
hundred clouds of gas and dust forming a disk-like structure ({\it
Circumnuclear Disk CND}, see e.g. G\"{u}sten et al. 1987) up to a
radius of $\sim$7 pc\footnote{We assume 8.5 kpc for the distance
to the Galactic Center.}. The CND has a hydrogen mass of a few
times 10$^{4}$~M$_{\odot}$ which is distributed in clumps with an
estimated area filling factor of $\phi_{\rm A} \sim 0.1$ and a
volume filling factor of $\phi_{\rm V} \sim 0.01$. The clumps have
densities of several 10$^{5}$~cm$^{-3}$, radii of $\sim 0.05$~pc
and gas temperatures $\geq$100~K. A typical dusty clump has a
visual extinction A$_{\rm V} > 30^{\rm m}$ and M$_{\rm H}
\sim$30~M$_{\odot}$. Further properties of the clouds in the
central 2~pc are listed in Jackson et al. (1993). The vertical
thickness ($2\,H$) of the CND increases from $\sim$0.5~pc at a
radius of 2~pc to about 2~pc at the outer radius of 7~pc. The disk
rotates with a velocity of $\sim$100 km\,s$^{-1}$ which
corresponds to a Keplerian velocity around a central object of
several 10$^{6}$~M$_{\odot}$ and has a velocity dispersion of
$\sim$ 20~km\,s$^{-1}$. With an $R/H$ of $\sim 5-7$ ($R$
being the radius to the center), the CND can be qualified as a
thin disk. At a distance of $\sim 1.5$~pc the molecular gas
density drops by more than a magnitude, i.e. the central 3~pc of
the Galaxy are void of molecular gas. In this sense the CND
resembles more a ring or a torus.

In active galactic nuclei (AGN) the
central luminous and compact object (e.g. the inner accretion
disk around the black hole) is likely to be obscured by the torus,
if the mean free path of clouds along the line of sight is greater
than one. This qualitatively obvious statement was
quantified by Nenkova et al. (2002) based on the comparison
between models of cloudy dust emission and IR spectral energy
distributions. They conclude that the number of clouds along the
line of sight should be 5--10, which corresponds to the mean free
path of clouds being 5--10 times smaller than the radius of the torus. 
Krolik \& Begelman (1988) were among the first to discuss the overall
physical properties of the circumnuclear gas in active galactic centers.
They already emphasized the importance of the cloudy structure of
the circumnuclear material.

Based on the unified model (Antonucci 1993) the spectroscopic
difference between Seyfert 1 and 2 nuclei is due to the line of
sight passing through the torus (Sy 2) or not (Sy 1). From the
relative numbers of Sy~1 and Sy~2, Schmitt et al. (2001)
derive a half-opening angle of $48^\circ$ for the obscuring torus.
But Tran (2003) shows in a recent paper that about half the Sy~2
nuclei derived from a 12$\mu$m sample do not have detectable
hidden broad line regions (HBLR) in spectropolarimetric
observations. Because the obscuration determined from X-ray
spectra is not systematically different for Sy~2 with HBLR and 
those without HBLR, it is
not obvious that the obscuring material in non-HBLR Sy2 nuclei,
which seem to be intrinsically fainter, is in form of a torus. It
is therefore difficult to determine the half-opening angle of the
torus and its relative thickness statistically.

More evidence can be found by imaging extended NLR in O{\sc iii}
for example. For some sources the extended NLR take the shape of
an ionization cone either due to a anisotropic radiation source or
a collimating torus (for a review see Wilson 1996, and
Mulchaey, Wilson \& Tsvetanov 1996 for a list of morphologically
identified sources). The latter scenario is supported by evidence
for obscuration at the apex of some of the observed cones. If this
torus is responsible for an ionization cone, it extends radially 
from a few pc to several tens of pc and its flaring should be 
$\alpha=(180^{\rm o}-\psi)/2$, where $\psi$ is the opening angle 
of the ionization cone. 
With a mean opening angle of $\bar{\psi} = 80^\circ \pm
20^\circ$, the disk flaring angle is $\alpha \sim 50^\circ \pm
10^\circ$ or $H/R \sim 1$, which means that these tori are
geometrically thick.

The CND in the Galactic Center in its current form is clearly not an
obscuring torus, because the mean free path of the clouds along the
radius is $l_{\rm cl} \sim 10$~pc and thus $R/l_{\rm cl} < 1$,
where $R = 7$~pc is the extent of the CND, i.e. the line of sight
through the torus. It is not a thick disk either, because $H/R \sim
0.2$.

In this article we investigate if there are common properties between
the CND in the Galactic Center and obscuring tori in Seyfert galaxies.
In Sect.~\ref{sec:theory} we recall the theory of clumpy accretion
disks. The equations needed for the comparison between the different
models and observations are given in Sect.~\ref{sec:comparison}.
The models are than applied to the CND and a thick, obscuring torus
(Sect.~\ref{sec:application}). We discuss the results in
Sect.~\ref{sec:discussion} and give our conclusions in
Sect.~\ref{sec:conclusions}.

\section{The theory of clumpy gas disks \label{sec:theory}}

\subsection{Turbulent disks}

In a recent work (Vollmer \& Beckert 2002, Paper I) we developed an analytical
model for clumpy accretion disks and included a simplified description
of turbulence in the disk. In contrast to classical accretion disk theory
(see, e.g., Pringle 1981),
we do not use the  ``thermostat'' mechanism, which implies a direct
coupling between the heat produced by viscous friction and the
viscosity itself which is assumed to be proportional
to the thermal sound speed. Thus, the viscosity, which is responsible
for the gas heating, depends itself on the gas temperature.
This leads to an equilibrium corresponding to a thermostat mechanism.
Instead, we used an energy flux conservation, where the potential
energy that is gained through mass accretion and differential rotation
is cascaded by turbulence from large to small spatial scales. The inertial
range of the turbulent cascade under consideration will not reach down to
the scale of microscopic dissipation. Instead we regard the energy to be
dissipated at an inner scale where the first cascade coming from
the largest scales ends or breaks.

In a first approximation this happens when the gas clouds become
self-gravitating and dissipation then takes place inside individual clouds.
Because the energy reservoir is the gravitational binding
energy in the potential of the galaxy, a star cluster, or a central black hole,
and the dissipation scale is set by self-gravity, we will call this scenario
the fully gravitational model (FG model) in the following. Turbulence
transfers the energy from the driving wavelength $l_{\rm driv}$ to the
dissipation wavelength $l_{\rm diss}$, which corresponds to the length
scale of a self-gravitating cloud.
The two length scales are linked by the scaling parameter $\zeta$ which
measures the extent of the inertial range.
For a Kolmogorov-like turbulent energy spectrum
$\zeta = (l_{\rm driv}/l_{\rm diss})^{\frac{3}{4}}$.

An additional 'free' parameter of the model is the Toomre-$Q$ parameter:
%------------------------
\begin{equation}
  Q = \frac{v_{\rm turb}\,\Omega}{\pi\,G\,\Sigma} \geq 1\ ,
  \label{eq:tq}
\end{equation}
%------------------------
where $v_{\rm turb}$ is the turbulent velocity, $\Omega$ the
angular velocity, $G$ the gravitational constant, and $\Sigma$ the
gas surface density of the disk. For all models we assume
$Q$ to be constant throughout the disk. If we can approximate the
total gas mass within a radius $R$ by $M_{\rm gas} = \pi R^2
\Sigma$ Toomre-$Q$ can be rewritten
%------------------------
\begin{equation}
  Q=\frac{v_{\rm turb}}{v_{\rm rot}}\frac{M_{\rm tot}}{M_{\rm gas}}\ ,
  \label{eq:tqint}
\end{equation}
%------------------------
where $M_{\rm tot}$ is the total enclosed mass and $v_{\rm rot}$
the rotation velocity.

Like in a standard accretion disk scenario a viscosity allows redistribution
of angular momentum and mass accretion together with a gain of kinetic energy
from the potential well in which the disk resides.
For the viscosity we use the following prescription:
%------------------------
\begin{equation}
  \nu=\zeta^{-1} v_{\rm turb} l_{\rm driv}\ .
  \label{eq:nuu}
\end{equation}
%------------------------
A set of equations is obtained with 3 parameters:
the Toomre parameter $Q$, the scaling parameter $\zeta$, and
the mass accretion rate within the disk $\dot{M}$.
These equations can be solved analytically and the results were
already used in Paper I to describe properties of our Galaxy's extended
gas disk. Besides the parameters above the solutions depend
on the $v_{\rm rot}$, and $R$.
Within the model it turned out that the driving wavelength
in Eq.(\ref{eq:nuu}) equals the disk height $l_{\rm driv}=H$.

In a second step we included the energy input due to supernova
(SN) explosions (Vollmer \& Beckert 2003, Paper II). Here the
energy flux provided by SNe is transfered by turbulence to smaller
scales where it is again dissipated. The SN energy flux is assumed
to be proportional to the local star formation rate
$\dot{\rho}_{*}$, which, in turn, is taken to be
proportional to the mean density and inversely proportional to the
local free fall time of clouds. These clouds have sizes that are a
factor $\delta$ smaller than the driving length scale. The factor
of proportionality is the probability to find a self-gravitating
cloud, i.e. the volume filling factor. Because the model works
with vertically integrated quantities, like the surface density,
the integration length in the vertical direction is assumed to be
the turbulent driving length scale, i.e. the vertical length scale
over which clouds can become self-gravitating:
$\dot{\Sigma}_{*}=\dot{\rho}_{*}\,l_{\rm driv}$. The SN energy per
unit time $\dot{E}_{\rm SN}$ per area $\Delta A$ is therefore
proportional to the local star formation rate $\dot{\Sigma}_{*}$:
%------------------------
\begin{equation}
  \frac{\dot{E}_{\rm SN}}{\Delta A} = \xi \dot{\Sigma}_{*}\ ,
  \label{eq:sfrsn}
\end{equation}
%------------------------
where the factor of proportionality $\xi$ is independent of the radius
in the disk.
Its normalization with Galactic values yields $\xi =
4.6\,10^{-8}$~(pc/yr)$^{2}$.

In the FG model it is assumed that the energy flux transported through
the turbulent cascade is supplied by mass accretion and the energy balance is
%------------------------
\begin{equation}
  \rho \nu \frac{v_{\rm turb}^{2}}{l_{\rm driv}}=-\frac{1}{2\pi}\dot{M}
  v_{\rm rot} \frac{\partial \Omega}{\partial R}\ .
  \label{eq:efc}
\end{equation}
%------------------------
In the case of SN driven turbulence the energy flux is determined
by the star formation rate
%------------------------
\begin{equation}
  \rho \nu \frac{v_{\rm turb}^{2}}{l_{\rm driv}}=\xi\,\dot{\Sigma}_{*}\ .
  \label{eq:sgzefc}
\end{equation}
%------------------------
In most gas disks on galactic scales the energy supply by supernovae dominates
the gravitationally released energy for the same disk properties except for
large $Q$ values. The situation changes if the disks are so thin, that SN bubbles
break out and most of their energy is lost from the disk.

Furthermore, we take into account that the clouds are formed at
the compressed edges during the interaction between SN remnants.
The size of the clouds $l_{\rm cl}$ is smaller than the
turbulent driving wavelength and we use $\delta=l_{\rm
driv}/l_{\rm cl}$ with $\delta \geq 1$.

%----------------------------------------------------------------------------------
\subsection{Collisional disks \label{sec:colldisk}}
While the turbulent disk model describes the ISM as a continuous
medium where a turbulent cascade can develop, this
changes when the self-gravitating clouds become stable. In such a
medium an equilibrium disk can form, if there are fragmenting or
partially elastic collisions (the clouds are magnetized) between
the clouds. These collisions give rise to a viscosity which
depends on the velocity dispersion of the clouds (we will call it
$v_{\rm turb}$ nonetheless) and the mean free path of cloud with
respect to cloud collisions $l_{\rm coll}$. If the collisional
timescale $t_{\rm coll} = l_{\rm coll}/v_{\rm turb}$ is larger or
equal to the dynamical timescale, the resulting viscosity (Pringle
1981) can be written as
%------------------------
\begin{equation}
  \nu = \tau^{-1}v_{\rm turb} H\ ,
\end{equation}
%------------------------
where $\tau = t_{\rm coll}\Omega$ and $H$ is the disk height. Here it 
is already assumed that the disk is in hydrostatic equilibrium vertically
$H = v_{\rm turb}/\Omega$, where the pressure is provided by the velocity
dispersion. The energy dissipation rate due to collisions can be written as
%------------------------
\begin{equation} \label{eq:en:coll}
  \frac{\Delta E}{\Delta A\,\Delta t}
  = \epsilon \frac{\frac{3}{2}\Sigma v_{\rm turb}^{2}}{t_{\rm coll}}
  = \epsilon \frac{\frac{3}{2}\Sigma v_{\rm turb}^{3}}{l_{\rm coll}}
%  =\epsilon \frac{\frac{3}{2}\Sigma v_{\rm turb}^{3}}{\tau H}
  = \Sigma \nu \left(R \frac{\partial \Omega}{\partial R}\right)^2 .
\end{equation}
%------------------------
The last equation on the right implies that the energy source is
differential rotation and accretion as in the FG model. Here
we introduced an efficiency parameter $\epsilon$ which
measures the fraction of the kinetic energy of clouds dissipated
in a typical collision. Because most conclusions from this model
are independent of the particular form of the energy equation
(\ref{eq:en:coll}) we set $\epsilon = 2/3$ and so we are
consistent with Eq. (\ref{eq:efc}) and the use in  Paper~I. Since
the driving wavelength in Eq. (\ref{eq:nuu}) for the FG model is
the disk scale height $l_{\rm driv}=H$, the FG model and the
collisional model are formally equivalent when we identify $\tau$
with $\zeta$. But the interpretation of the viscosity in terms of
a continuous turbulent medium (FG model) and the collisional
viscosity here are completely different.

%___________________________________________________________________________________
\section{A Comparison of characteristic properties of the three models
\label{sec:comparison}}

In this section we recall relevant relations between observables within the FG
and the SN models and derive observable properties of the collisional disk.
This will allow us later on to determine the free model parameters and compare
the models with observations.

\subsection{The fully gravitational (FG) model \label{FGBasics}}
Following Paper~I this model has two dimensionless parameters:
Toomre-$Q$ defined in Eq. (\ref{eq:tqint}) and $\zeta$ in
Eq.~(\ref{eq:nuu}). From a solution consistent with
observations it is possible to derive the mass accretion rate in
the disk $\dot{M}$ from the FG model. Very importantly the
turbulent velocity dispersion can be expressed as a function of
the gravitational constant $G$, the scaling parameter $\zeta$, the
mass accretion rate $\dot{M}$, and $Q$:
%------------------------
\begin{equation}
  v_{\rm turb}=\big(\frac{1}{2}G \zeta Q \dot{M} \big)^{\frac{1}{3}}\ .
  \label{eq:vturbfg}
\end{equation}
%------------------------
Furthermore the cloud radius is given by
%------------------------
\begin{equation}
  r_{\rm cl}=\zeta^{-1}H\ ,
  \label{eq:rclfg}
\end{equation}
%------------------------
and the volume filling factor, which is defined by
$\phi_{\rm V}=\rho/\rho_{\rm cl}$, where $\rho$ is the mean density
in the disk and $\rho_{\rm cl}$ is the density of the clouds, yields:
%------------------------
\begin{equation}
  \phi_{\rm V}=\frac{1}{\zeta Q}\ .
  \label{eq:phiFG}
\end{equation}
%------------------------
For a disk with almost constant surface density the total gas mass is given by
%------------------------
\begin{equation}
  M_{\rm gas} = \pi \Sigma R^{2} =
  2^{-\frac{1}{3}}G^{-\frac{2}{3}}\zeta^{\frac{1}{3}}
  Q^{-\frac{2}{3}}\dot{M}^{\frac{1}{3}}v_{\rm rot}R
  \label{eq:mgas}
\end{equation}
%------------------------
where the rotational velocity $v_{\rm rot}$ can be either directly observed
or determined from the total mass distribution. In the case of an isothermal
star cluster the surface density also has a cusp with $\Sigma \propto R^{-1}$
and the enclosed gas mass is a factor 2 larger than in Eq.(\ref{eq:mgas}).

%_________________________________________________________________________________
\subsection{The SN model}

In Paper II we distinguish two cases: (i) a self-gravitating gas
disk in vertical direction ($Q=1$) and (ii) $Q>1$.

For $Q=1$ the velocity dispersion in the disk is given by
%------------------------
\begin{equation} \label{eq:vturbQ1}
  v_{\rm turb}=0.82\,G^{\frac{3}{11}} \dot{M}^{\frac{3}{11}}
  \delta^{-\frac{1}{11}} \xi^{\frac{1}{11}}\ ,
\end{equation}
%------------------------
where $\xi$ is the normalization of the SN energy input in
Eq. (\ref{eq:sgzefc}) and $\delta$ the ratio if actual cloud size to
the driving length scale.

For $Q>1$ the velocity dispersion in the disk and the driving length
scale are
%------------------------
\begin{eqnarray}
  v_{\rm turb} & = & 0.87\,\chi^{\frac{4}{15}} G^{\frac{1}{5}}
  \dot{M}^{\frac{1}{5}} Q^{\frac{4}{15}} \delta^{-\frac{1}{15}}
  \xi^{\frac{1}{15}} v_{\rm rot}^{\frac{4}{15}}\ ,
  \label{eq:vturbsn} \\
  l_{\rm driv} & = & 0.69\,\chi^{-\frac{1}{6}}G^{\frac{1}{2}}
  \dot{M}^{\frac{1}{2}}Q^{\frac{5}{6}}\delta^{\frac{1}{6}}
  \xi^{-\frac{1}{6}} v_{\rm rot}^{-\frac{7}{6}} R\ .
  \label{eq:ldrivsn}
\end{eqnarray}
%------------------------
where $\chi=H/R$ is the relative thickness of the disk.
The volume filling factor turns out to be
%------------------------
\begin{equation}
  \phi_{\rm V}=0.79\,\chi^{-\frac{2}{15}}G^{\frac{2}{5}}
  \dot{M}^{\frac{2}{5}}Q^{-\frac{2}{15}}\delta^{-\frac{22}{15}}
  \xi^{-\frac{8}{15}} v_{\rm rot}^{-\frac{2}{15}}\ .
  \label{eq:phivsn}
\end{equation}
%------------------------

It is possible to give a criterion for the validity of the model that is
when the energy input through supernovae explosions Eq. (\ref{eq:sgzefc})
dominates over the gravitationally released energy due to accretion
(Eq. \ref{eq:efc}). This can be turned into the following condition for
the volume filling factor:
%------------------------
\begin{equation} \label{rel:sn}
  \phi_{\rm V} > \xi^{-2} \Omega^{-4} v_{\rm turb}^{8}
  l_{\rm driv}^{-4}\ .
\end{equation}
%------------------------

\subsection{The collisional model \label{sec:collmodel}}
For the collisional model we can get quite general relations for cloud sizes
and a criterion for the $Q$ parameter to be satisfied so that clouds are
actually stable. As already mentioned in Sect.~\ref{sec:colldisk},
the mean free path of a cloud is
%------------------------
\begin{equation}
l_{\rm coll}=\frac{v_{\rm turb}}{t_{\rm coll}}=\tau \frac{v_{\rm
turb}}{\Omega}=\tau H\ ,
\end{equation}
%------------------------
while the mean free path on the other hand is the inverse of the cloud number
density $n$ times the cloud cross section $\sigma= \pi r_{\rm cl}^{2}$:
%------------------------
\begin{equation}
  l_{\rm coll}=(n \sigma)^{-1}=\big(\frac{3\,\Sigma \phi_{\rm V}}
  {4\,r_{\rm cl} \rho H}\big)^{-1}=
  \frac{4\,r_{\rm cl}}{3\,\phi_{\rm V}}\ .
\end{equation}
%------------------------
In contrast to the turbulent models, the condition for clouds to be
self-gravitating is that the free fall time equals the sound crossing
time (in the case of the turbulent models it is the turbulent crossing time):
%------------------------
\begin{equation} \label{eq:ff}
  t_{\rm ff}=\sqrt{\frac{3\,\pi}{32\,G\rho_{\rm cl}}}
  =t_{\rm sound}=\frac{r_{\rm cl}}{c_{\rm s}}\ ,
\end{equation}
%------------------------
where $c_{\rm s}$ is the sound speed within the clouds.
Inserting $\rho=\Omega^{2}/(\pi G Q)$ from the definition of the $Q$-parameter
(Paper I) leads to
%------------------------
\begin{equation}
  r_{\rm cl}=\frac{\pi^2}{8} \frac{Q}{\tau} \Omega^{-1}
  v_{\rm turb}^{-1}c_{\rm s}^{2}\ .
  \label{eq:rcl}
\end{equation}
%------------------------
Not surprisingly the size of clouds gets larger when the disk is further 
away from the gravitational instability (larger $Q$ values) and the collision 
time is long. The critical density against tidal shear is
$\rho_{\rm crit}=3 \Omega^{2}/(4 \pi G)$ and thus 
$\rho/\rho_{\rm crit}=4/(3 Q)$.
On the other hand we have $\rho/\rho_{\rm cl}=\phi_{\rm V}$. The condition for
the clouds to resist disruption by tidal shear is $\rho_{\rm cl} >
\rho_{\rm crit}$, which leads with the help of Eq. (\ref{eq:ff}) to an upper
bound for the cloud size:
%------------------------
\begin{equation}
  r_{\rm cl} < \frac{\pi}{\sqrt{8}}\frac{c_{\rm s}}{\Omega}\ .
\end{equation}
%------------------------
Inserting Eq.~(\ref{eq:rcl}) for the cloud size gives the following condition
for the ratio of Toomre parameter to the dimensionless collision time $\tau$
%------------------------
\begin{equation}
  \frac{Q}{\tau} <  \frac{\sqrt{8}}{\pi} \frac{v_{\rm turb}}{c_{\rm s}}
  \label{eq:qcondition}
\end{equation}
%------------------------
or using the definition of the Toomre-$Q$ (Eq. \ref{eq:tq}) a lower bound on
the collision time for stability of clouds
%------------------------
\begin{equation}
  \tau >  \frac{\Omega c_s}{\sqrt{8} G \Sigma}\ .
  \label{eq:tcondition}
\end{equation}
%------------------------
The upper bound on $Q/\tau$ increases for thick disks
(e.g. large velocity dispersions), while the
limit on the collision time gets shorter for more massive disks.

The limit on $Q/\tau$ is especially interesting for geometrically thick and
obscuring tori, because obscuration requires $\tau = l_{\rm coll}/H$ to be small,
which turns into a severe constrain on the allowed $Q$-values.

For the CND the clumpy structure is allowed to be transparent, but
with the known cloud sizes and temperatures (sound speed)
it is possible to get a measure of $\tau$ from the formally
equivalent FG model expressions Eq.~(\ref{eq:vturbfg},
\ref{eq:mgas}) by replacing $\zeta$ by $\tau$, and
Eq.~(\ref{eq:rcl})
%------------------------
\begin{equation}
  \tau = \frac{\pi^2}{8} \frac{c_{\rm s}^{2}}{ G M_{\rm gas}}
  \frac{R^{2}}{r_{\rm cl}} \, .
  \label{eq:taucoll}
\end{equation}
%------------------------
Obviously the collision time gets longer for smaller clouds, less massive disk,
but surprisingly also for higher temperatures in the clouds.

%------------------------------------------------------------------------------------
\section{Finding the most appropriate model \label{sec:application}}

In this Section we will determine the model parameters of the 3 different
models for (i) the Circumnuclear Disk in the Galactic Center and (ii)
for obscuring tori. This will enable us to discuss, which models are
realistic for the two kinds of ring structures.

\subsection{What it needs to make a Circumnuclear Disk}

The observed or from observations deduced properties of the CND are summarized in
Tab.~\ref{tab:cndtable}.
%------------------------
\begin{table}
      \caption{Observed properties for the CND}
         \label{tab:cndtable}
      \[
       \begin{array}{|l|l|ll|}
        \hline
        {\rm cloud\ radius} & r_{\rm cl} & 0.05 & {\rm pc} \\
 \hline
 {\rm cloud\ mass} & M_{\rm cl} & 30 & {\rm M}_{\odot} \\
  \hline
 {\rm sound\ speed\ within\ the\ clouds} & c_{\rm s} & 1.05 & {\rm
km\,s}^{-1} \\
 \hline
 {\rm turbulent\ velocity\ dispersion\ in\ the\ disk} & v_{\rm turb} &
20 & {\rm km\,s}^{-1} \\
 \hline
 {\rm disk\ radius} & R & 7 & {\rm pc} \\
 \hline
 {\rm disk\ height} & H & 1 & {\rm pc} \\
 \hline
 {\rm disk\ mass} & M_{\rm gas} & 2\,10^{4} & {\rm M}_{\odot} \\
 \hline
 {\rm rotation\ velocity} & v_{\rm rot} & 115 & {\rm km\,s}^{-1} \\
 \hline
        \end{array}
      \]
\end{table}
%------------------------
The CND resides in a region where the central star cluster starts to dominate
the total mass (e.g. Sch\"{o}del et al. 2002).
The total enclosed mass at radius $R$ rises linearly with
$R$ and the rotation velocity is almost constant throughout the CND.
Because the CND extends only over a limited range in radius we use
$M_{\rm gas} = \pi \Sigma R^{2}$ for the gas mass contrary to the discussion at
the end of Sec.~\ref{FGBasics}.
For all models the Toomre-$Q$ parameter is
%------------------------
\begin{equation} \label{qobs:CND}
  Q=\frac{v_{\rm turb}v_{\rm rot}R}{G M_{\rm gas}}=190\ .
\end{equation}
%------------------------
With this number fixed we can go to the individual models

\subsubsection{Massive clouds in the FG model}
From the cloud size and the disk height inserted in
Eq.~(\ref{eq:rclfg}) we get $\zeta=20$. Together with the $Q$-value
this leads to  a volume filling factor $\phi_{\rm V}=(\zeta Q)^{-1}
\approx 2.6\,10^{-4}$.
Independent of the volume filling factor we find from Eq.~(\ref{eq:vturbfg})
a mass accretion rate of $\dot{M} = 10^{-3}$~M$_{\odot}$yr$^{-1}$.
The mass of a single cloud follows
$M_{\rm cl}=(4/3)\,\pi r_{\rm cl}^{3} \rho / \phi_{\rm V}= 260 $~M$_{\odot}$,
but notice that this a rather strong function of the assumed outer radius
of the CND, because it enters quadratically in the mass density.
The number of individual clouds in the CND is only 80 in FG model.

\subsubsection{Why supernovae do not play any role in the Circumnuclear Disk
\label{Sec:CND:SN}}

In a first attempt we followed Paper II and assumed $\delta=5$,
but the resulting driving length scale was much to large. We could
reduce that problem by choosing $\delta=2$. The relative
thickness of the disk is known $\chi=H/R=0.14$, so that we derived
from Eq.~(\ref{eq:vturbsn}) a mass accretion rate of
$\dot{M}=2.3\,10^{-3}$~M$_{\odot}$yr$^{-1}$. The expression for
the volume filling factor (Eq.~\ref{eq:phivsn}), the driving
length scale (Eq.~\ref{eq:ldrivsn}), and the cloud radius then
give $\phi_{\rm V}=8.2\,10^{-4}$, $l_{\rm driv}=1.24$~pc, and
$r_{\rm cl} =l_{\rm driv}/\delta=0.62$~pc. 
In this scenario the cloud radius is derived from the model and 
is listed in Table~\ref{tab:resume} for that reason. 
Consequently, the mass
of a single gas cloud is $M_{\rm cl}=1.6\,10^{5}$~M$_{\odot}$ even
larger than the whole mass of the CND. 

In a further approach we treated $\delta$ as a free parameter, solving
Eq.~(\ref{eq:ldrivsn} and \ref{eq:vturbsn}) together
with $r_{\rm cl}=l_{\rm driv}/\delta$  leads to the incredible ratio
%------------------------
\begin{equation}
  \delta =0.86 Q \zeta^{-5} \frac{R}{r_{\rm cl}}
\frac{v_{\rm turb}^{15}}{v_{\rm rot}^{11} \xi^{2}} = 7.4\,10^6\, ,
\end{equation}
%------------------------
which states that $l_{\rm driv} \approx 370$~kpc.

In a last attempt we left $\xi$ the normalization of the SN energy input rate
as a free parameter. This requires us to fix the driving length scale
$l_{\rm driv}$, which up to now was always too large, to the disk scale height
$l_{\rm driv} = H = 1$~pc. The cloud radius from Table~\ref{tab:cndtable} 
is used as input here and is included in Table~\ref{tab:resume} 
for completeness only. This immediately leads to $\delta = 20$,
an accretion rate of $\dot{M}=6\,10^{-3}$~M$_{\odot}$yr$^{-1}$,
a volume filling factor of $\phi_{\rm V}= 10^{-6}$, and a cloud mass of
$M_{\rm cl}=5\,10^{4}$~M$_{\odot}$. Apart from the much too large cloud mass
one sees that this model also fails for the SN efficiency, because we find
$\xi = 2.7\,10^7$~(pc/yr)$^2$, which requires star formation and the
resulting SN energy input into the disk to be about 600 times more effective
than elsewhere in the Galaxy. The large driving length scale or the large star
formation/SN efficiency required shows, that energy supplied by SN under
normal conditions is much too small to drive the dynamics of the CND.

\subsubsection{A transparent collisional CND \label{CollCND}}

The observed properties of the CND allow us to estimate the 'transparency'
$\tau = l_{\rm{coll}}/H$ of the CND from Eq.~(\ref{eq:taucoll}) which results
in $\tau=15$. So on average there are about $\tau/(2 \pi) \approx 2$
collisions for each cloud per orbit in this model. The expression for the
velocity dispersion Eq.~(\ref{eq:vturbfg}), which is equally valid for the
collisional case with $\zeta = \tau$, gives an estimate of the mass accretion
rate of $\dot{M}=1.3\,10^{-3}$~M$_{\odot}$yr$^{-1}$, which is almost the same
as in the FG model, because $\tau$ here and $\zeta$ for in the FG case are not
much different. The volume filling factor of clouds
$\phi_{\rm V}=(4\,r_{\rm cl} \Omega)/(3\,\tau v_{\rm turb})=3.6\,10^{-3}$
is significantly larger than in the FG model. The reason is that $Q$ enters
in $\phi_{\rm V}$ of the FG model (Eq. \ref{eq:phiFG}), while it is the cloud
size in the collisional case here. The result is a substantially smaller mass
of typical clouds $M_{\rm cl}=19$~M$_{\odot}$. While the transparency derived
from Eq.~(\ref{eq:taucoll}) depends on global properties like the total gas
mass in the CND, $\tau$ can also be determined from a purely geometrical
consideration
%------------------------
\begin{equation}
  \tau=\frac{\Omega}{n \sigma v_{\rm turb}} = 20\ ,
\end{equation}
%------------------------
with the cloud number density $n=N_{\rm cl}/V=(M_{\rm gas}/M_{\rm cl})/V$,
the total number of clouds $N_{\rm cl}=1044$, and $V = 200$~pc$^{3}$ the
volume of the disk. This value is reasonably close to the one derived
from Eq.~(\ref{eq:taucoll}) above.

\subsection{Building Geometrically and Optically Thick Tori \label{sec:thicktori}}
For obscuring the BLRs in type 2 AGN and for collimating outflows and the
ionizing flux in the extended NLR (ionization cones) we suggest properties
for an obscuring torus as summarized in Tab.~\ref{tab:torustable}. The
values would be appropriate for our galactic center becoming a Sy~2 nucleus
with a compact dusty torus. The results are easily rescaled to objects
like NGC~1068, where NIR-speckel observations (Wittkowski et al. 1998) show
that the inner rim of the torus, if interpreted as such, is less than
2~pc from the AGN.
%------------------------
\begin{table}
      \caption{Assumed properties of an obscuring torus}
         \label{tab:torustable}
      \[
       \begin{array}{|l|l|ll|}
        \hline
 {\rm sound\ speed\ within\ the\ clouds} & c_{\rm s} & 1.2 & {\rm
km\,s}^{-1} \\
 \hline
 {\rm turbulent\ velocity\ dispersion\ in\ the\ disk} & v_{\rm turb} &
72 & {\rm km\,s}^{-1} \\
 \hline
 {\rm disk\ radius} & R & 10 & {\rm pc} \\
 \hline
 {\rm disk\ height} & H & 6 & {\rm pc} \\
 \hline
 {\rm rotation\ velocity} & v_{\rm rot} & 120 & {\rm km\,s}^{-1} \\
 \hline
        \end{array}
      \]
\end{table}
%------------------------
The torus rotates somewhat faster and since there is an enhanced
radiation from the active nucleus, the sound speed within the clouds
is somewhat higher.

As already mentioned in Sect.~\ref{sec:intro} the torus is obscuring if the
mean free path of the clouds $l_{\rm coll}$ is smaller than the line of
sight through the torus. In a first approximation this corresponds to
the condition $l_{\rm coll}=H$ where we expect that in the midplane the
line of sight is blocked by several clouds (Nenkova et al. 2002) and
the torus is also geometrically thick $H\sim R$.
Within the framework of the FG and collisional models this leads
immediately to the requirement $\zeta=\tau=1$.

We can start the discussion with the statement that the FG model 
fails to describe a thick torus, because
the requirement $\zeta=1$ implies that the cloud radius equals the disk 
height $r_{\rm cl}=H/\zeta=H$.
Since this is unphysical, we can reject this model for obscuring tori.

\subsubsection{Colliding clouds in a thick torus \label{Sec:coll_torus}}
From the condition that the gas clouds have to resist tidal shear
Eq.~(\ref{eq:qcondition}) reads $Q \le 54$. As the torus is
thicker for larger $Q$ we assume $Q$ to take its maximal allowed
value. This means that typical clouds are actually at the shear
limit. Together with $\tau=1$ we find clouds of $r_{\rm
cl}=0.1$~pc. From Eq.~(\ref{eq:vturbfg}) like in the CND case we
find an accretion rate of $\dot{M}=3.3$~M$_{\odot}$yr$^{-1}$ and
from (Eq.~(\ref{eq:mgas}) a gas mass in the torus of $M_{\rm
gas}=4\,10^{5}$~M$_{\odot}$. The volume filling factor is
determined in analogy to the CND case (Sec.~\ref{CollCND}) and gives
$\phi_{\rm V}= 2.5\,10^{-2}$. The resulting mass of a single gas
cloud is here $M_{\rm cl}=45$~M$_{\odot}$. 
The total gas mass is larger than 
in the collisional CND case as expected, because more mass in clouds is 
needed for obscuration. The implied cloud sizes and cloud masses are
surprisingly similar to those of the CND, which may indicate that the 
dominating physical processes in CND and torus are the same. The volume 
filling factor is of course larger as we would not get an obscuring 
torus otherwise. The higher density of clouds now lead to more cloud 
collisions and consequently the increased mass accretion rate, 
which feeds the AGN.

\subsubsection{Supernovae can drive tori at large mass accretion rates}
For the SN driven turbulent case of a torus model it is not obvious
from the beginning what the appropriate $Q$-parameter is. The torus
will only be geometrically thick for large turbulent velocities. This
together with the normalization of the SN efficiency $\xi$ derived
from integrated galactic values (Paper II), implies that a mass accretion
rate (Eq.~\ref{eq:vturbQ1}) of $\dot{M} \sim 45\, \delta^{1/3}$~M$_{\odot}$/yr
would be needed for a ($Q$=1)-torus.

In the case of a torus with $Q>1$ we can estimate, if the
energy supplied by supernovae will dominate the heating rate due
to viscosity and differential rotation.
%(Eq.~\ref{eq:sndominate}).
Inserting the volume filling factor from Eq. (\ref{eq:phivsn})
into Eq. (\ref{rel:sn})
leads to a limit for the product $ \delta Q^2$:
%------------------------
\begin{equation}
\delta Q^2 \leq \frac{\zeta^{3} \xi}{\chi^{2} v_{\rm rot}^{2}}\ .
\label{eq:deltacondition}
\end{equation}
%------------------------
With the values for the torus model from Table
\ref{tab:torustable} it gives a limit $\delta Q^2 < 8.6\,
\zeta^3$. The comparison in the inequality
(\ref{eq:deltacondition}) is made with the collisional case and
for a torus it implies $\zeta = \tau =1$. This results in a severe
limit for both $\delta$ and $Q$, because both parameters should be
larger than $1$. The range of allowed values for $\delta$ and $Q$ 
is rather small and we can assume $\delta = 2$ and $Q=2$
for the moment. The model is now completely fixed and we find a huge 
mass accretion rate of $\dot{M}= 42$~M$_{\odot}$yr$^{-1}$ 
through the torus, 
with cloud sizes of $r_{\rm cl}\sim 2$~pc and cloud masses of
$280$~M$_{\odot}$. The torus in this model has a volume filling
factor of clouds of $\phi_V \sim 6\,10^{-2}$ and a total mass in
gas of $M_{\rm gas} \sim 1\,10^{7}$~M$_{\odot}$. The large cloud
sizes and $\delta=2$ implies that the driving length scale is
$l_{\rm{driv}} = 0.7 H$ almost as large as the torus scale
height. Because $Q$ is small in this model and the torus is
required to be geometrically thick with $H \sim R$, the gas mass
in the torus is comparable to the total enclosed mass in stars and
central black hole.

%__________________________________________________________________________
%__________________________________________________________________________
\section{Prerequisites and Consequences \label{sec:discussion}}
In the previous section we have quantified what is needed for turbulent
(fully self-gravitational, or supernovae driven) models and a model
consisting of quasi-stable clouds undergoing collisions, which lead to
energy dissipation balanced by accretion in the gravitational potential
of the galactic center.
A r\'esum\'e of the results for the Circumnuclear Disk and an obscuring torus
is given in Tab.~\ref{tab:resume}

%------------------------
\begin{table}
      \caption{Derived parameters and properties of the CND and an
   obscuring torus (TORUS). The values in brackets for the SN model are for
   $\xi$  treated as a free parameter as described
   in Sec.~\protect\ref{Sec:CND:SN}. }
         \label{tab:resume}
      \[
       \begin{array}{|l|l|l|l|l|}
 \hline
  & & {\rm CND} & {\rm TORUS} & {\rm units}\\
        \hline
 \hline
 {\rm \bf FG\ model} & & & & \\
 \hline
 {\rm Toomre\ parameter} & Q & 190 & - & \\
 \hline
 l_{\rm driv}/l_{\rm diss}  & \zeta & 20 & 1 & \\
 \hline
 {\rm mass\ accretion\ rate} & \dot{M} & 1\,10^{-3} & - & {\rm
M}_{\odot}{\rm yr}^{-1} \\
 \hline
 {\rm volume\ filling\ factor} & \phi_{\rm V} & 2.6\,10^{-4} & - & \\
 \hline
 {\rm cloud\ mass} & M_{\rm cl} & 260 & - & {\rm M}_{\odot} \\
 \hline
 {\rm \bf SN\ model} & & & & \\
        \hline
 {\rm Toomre\ parameter} & Q & 190 & \sim 2 & \\
 \hline
 l_{\rm driv}/l_{\rm cl} & \delta & 2(20) & \sim 2 & \\
 \hline
 {\rm mass\ accretion\ rate} & \dot{M} & 2.3\,10^{-3}(6\,10^{-3}) & 42 &
{\rm M}_{\odot}{\rm yr}^{-1} \\
 \hline
 {\rm volume\ filling\ factor} & \phi_{\rm V} & 8\,10^{-4}(10^{-6})
& \sim 6\,10^{-2} & \\
 \hline
 {\rm cloud\ radius} & r_{\rm cl} & 0.62(0.05) & 2.0 & {\rm pc} \\
 \hline
 {\rm cloud\ mass} & M_{\rm cl} & 1.6\,10^{5}(5\,10^{4}) &  280
& {\rm M}_{\odot} \\
  \hline
 \hline
 {\rm \bf collisional\ model} & & & & \\
        \hline
 {\rm Toomre\ parameter} & Q & 190 & \sim 54 & \\
 \hline
 t_{\rm coll} \Omega & \tau & 15 & 1 & \\
 \hline
 {\rm mass\ accretion\ rate} & \dot{M} & 1.3\,10^{-3} & 3.3 & {\rm
M}_{\odot}{\rm yr}^{-1} \\
 \hline
 {\rm volume\ filling\ factor} & \phi_{\rm V} & 3.6\,10^{-3} & 
 2.5\,10^{-2} & \\
 \hline
 {\rm cloud\ mass} & M_{\rm cl} & 20 & 45 & {\rm M}_{\odot} \\
 \hline
 {\rm disk\ mass} & M_{\rm gas} & 2\,10^{4} & 4\,10^{5} & {\rm
M}_{\odot} \\
  \hline
  \hline
        \end{array}
      \]
\end{table}
%------------------------
%
For the Circumnuclear Disk (CND) in the Galactic Center the model
of supernovae driven turbulence (SN model) can clearly be
rejected, because the inferred cloud mass is larger than
$10^{4}$~M$_{\odot}$ and all the mass in the CND would be only one
cloud. Within the framework of the fully gravitational (FG) model,
the volume filling factor is small ($\phi_{\rm V}=
2.6\,10^{-4}$) and the cloud mass is about 8 times the
observed value ($M_{\rm cl}^{\rm obs} \sim
30$~M$_{\odot}$). Even when allowing for the
uncertainties in the observational parameters like total mass and
outer radius of the CND and the crudeness of the approximations in
the model, it is nonetheless unlikely that this model is a
realistic description of the CND. In contrast to the SN/FG-models
the collisional model leads to disk and cloud properties that are
close to the observed values. This model thus gives the best
approximation to the observations. It shares with the FG model the
property that accretion is the energy source for keeping the
velocity dispersion, but it consists of stable, individual clouds,
which are not part of a turbulent cascade.
This conclusion is in qualitative agreement with those
already drawn by Krolik \& Begelman (1988). These authors
derived the cloud properties on the basis of an extrapolated, observed 
gas pressure. They then used a model for an equilibrium configuration where cloud
mergers balance tidal shearing to show that the cloud area filling factor should be one. 
In this case the collisional time scale and the time scale for destruction by tidal shear
are comparable. In our collisional model, the cloud properties
are best described by the transparency $\tau$, which is not fixed to a singular value, 
and the fact that the clouds must resist tidal shear (Sec.~\ref{sec:thicktori}).

%__________________________________________________________________________
\subsection{The efficiency of star formation and SN energy input in
a collisional scenario}
In the following we compare the energy input due to SNe to the energy
dissipation rate per unit area due to collisions within the collisional
model described above. Supernovae can be important if their energy
supplied to the ISM dominates the dissipation rate
due to collisions. This is nessesary but not sufficient to influence 
the dynamics of clouds, as we show below based on the ratio of cloud to 
SN remnant (SNR) surface area. 

For the first step of the argument the energy supply rate from
Eq.~(\ref{eq:sfrsn}), modified by an efficiency $\eta$,
must be larger than the dissipation rate
due to collisions. The star formation rate in a collisional scenario
can be estimated following the discussion of Eq.~(19) in Paper II by
$\dot{\Sigma}_\star = \phi_V \Sigma/ t_{\mathrm{ff}}$. 
The free-fall time-scale is 
relevant for the size of clouds in the collisional case (Eq.~\ref{eq:ff}) 
and $r_{\mathrm{cl}}$ taken from Eq.~(\ref{eq:rcl}) with $Q$ at its 
maximum (Eq.~\ref{eq:qcondition}) leads to the relation  
%------------------------
\begin{equation}
   \eta \xi  \dot{\Sigma}_\star  = \frac{\sqrt{8}}{\pi} \eta \phi_V \xi 
   \Sigma \Omega > 
   \Sigma \frac{v_{\rm turb}^{3}}{\tau H}\ .
\end{equation}
%------------------------
From that we get a minimum efficiency $\eta_{\rm{eq}}$ for which SNe
and accretion supply equal amounts of energy to be dissipated:
%-----------------------------------------------
\begin{equation}
 \eta_{\rm{eq}} = \frac{\pi}{\sqrt{8}}
 \frac{v_{\rm turb}^{2}}{\tau \phi_V \xi} \,.
\end{equation}
%-----------------------------------------------
For the collisional torus model described in Sec.~\ref{Sec:coll_torus} we find
$\eta_{\rm{eq}} = 5.2$ so that star formation and SNe are required to be
more efficient than in galactic disks to become important. If the energy of 
SNe would be kept in the cloudy structure of the CND the 
equilibrium  value would be $\eta_{\rm{eq,CND}} = 0.13$, where we have used 
the `observed' value for $Q$ from Eq.~(\ref{qobs:CND}). We conclude
that if star formation in the Galactic Center proceeds in the same way as
in the Galactic disk, SN will provide a perhaps not negligible energy input
in addition to dissipative collisions.

Nevertheless, an equilibrium disk scenario where the dynamics of clouds is
driven by the energy input due to SNe is quite unlikely as we will show next.
The SNR will expand into the inter-cloud medium (ICM). The ICM
density and thermal velocity will limit the size the SNR and therefore the
sphere of influence of the SN. To get an estimate for the ICM density
we assume that it is in pressure balance with the clouds
and fills the same region of space as the clouds do, which implies
that the sound speed in the ICM equals the velocity dispersion
$v_{\rm{turb}}$. The ICM density is
therefore $\rho_{\rm{ICM}} = \rho_{\rm{Cl}} c_s^2/v_{\rm{turb}}^2$.
With the volume filling factor of the collisional model
%---------------------------
\begin{equation}
  \rho_{\rm{ICM}} = \frac{6}{\pi^2} \frac{\tau^2}{Q} \rho \, ,
\end{equation}
%---------------------------
where $\rho$ is the mean density in torus or CND. It turns out that the
ICM density is $70\%$ of the mean density for the CND and about 1\% for
the torus model; in both cases we find a particle density of
roughly $10^{3}$~cm$^{-3}$, which is nice agreement with the result of 
Erickson et al. (1994). We can now estimate the maximum SNR size by
using the power law fits provided by Thornton et al. (1998). To do that
one has to realize that the SNR resolves when the shock speed of the SNR
falls to the sound speed in the external medium which is the
velocity dispersion $v_{\rm{turb}}$ here. With $v_s \propto R_{\rm{SN}}^{-2}$
we get from extrapolating the Thornton et al. result
%---------------------------
\begin{equation} \label{eq:RSNR}
  R_{\rm{SN}}  =  1.1\, E_{51}^{2/7}\, n_3^{-0.42}\, (Z/Z_\odot)^{-0.1}
  \,\rm{pc}\, .
\end{equation}
%---------------------------
In the CND most SNR are likely to break out of the relatively thin disk
of molecular clouds and the efficiency will be largely reduced by this
effect. In the case of the torus the fraction of SN energy impinging
on clouds is simply the cross section of one cloud times the mean
number of clouds within the SNR of radius Eq.~(\ref{eq:RSNR}) divided
by the surface area of the SNR. This results in
$\eta = R_{\rm{SN}} \phi_V^2 /(4 r_{\rm{cl}})
= \phi_V R_{\rm{SN}}/(3 \tau H)$ and the efficiency of SNe energy from star 
formation to drive cloud motion
is $\eta = 1.5\,10^{-3}$ for the torus model.
Thus we conclude that both CND and obscuring torus can be successfully
described by the collisional model. The SN energy will almost exclusively
heat the inter-cloud medium and they will have little effect on the dynamics
of clouds in a scenario where clouds are stable and 'long-lived'.
Krolik \& Begelman (1988) used different arguments to estimate the
energy input due to SNe and stellar winds, which is necessary 
to maintain a thick torus. They also found that the stirring of the 
circumnuclear material due to stellar driven processes (supernovae and/or 
stellar winds) is always weaker than the effects of cloud collisions.

%__________________________________________________________________________
\subsection{The transition from a CND to an obscuring torus and vice versa}
From the discussion up to now it seems that both, the Circumnuclear Disk
in the Galactic Center and an obscuring torus, are well described by our
collisional model. The density of clouds within the model relative to
the critical density against tidal shear $\rho_{\rm crit}$ is
$\rho_{\rm cl}/\rho_{\rm crit}=4/(3 Q \phi_{\rm
V})$.
For the CND we find $\rho_{\rm cl}/\rho_{\rm crit}=1.8$ and the model
for the obscuring torus is constructed in a way so that $Q$ takes its
critical value from Eq.~(\ref{eq:qcondition}) which automatically means
that clouds are marginally stable against tidal shear.
Both clump densities are close to the critical density with
respect to tidal shear and we may speculate that both structures
are governed by the same physical processes.

In this sense the CND would be an obscuring torus if it had more mass
and it would be thick if it had a higher velocity dispersion.
We can quantify this with the definition of the (integrated)
Toomre parameter $Q$ (Eq.~\ref{eq:tqint}) to get
$ \frac{H}{R} = Q \frac{M_{\rm{gas}}}{M_{\rm{tot}}}$,
where $M_{\rm{tot}}$ is the total enclosed mass and $M_{\rm{gas}}$
the enclosed gas mass at that radius. If clouds are indeed at the shear limit
(Eq.~\ref{eq:qcondition}) we can replace $Q$ and get an expression for
the transparency
%-----------------------
\begin{equation}\label{eq:crittau}
 \tau = \frac{\pi}{\sqrt{8}} \frac{c_s}{v_{\rm{rot}}}
 \frac{M_{\rm{tot}}}{M_{\rm{gas}}}\, .
\end{equation}
%-----------------------
Because total mass and rotation velocity are given by the environment
in the galactic center and the sound speed within the clouds is almost
constant, we see that the torus becomes obscuring (that means $\tau \sim 1$)
for large gas masses in the torus. Inserting Eq.~(\ref{eq:crittau}) and
the corresponding expression for $Q$ in Eq.~(\ref{eq:vturbfg}) to
derive the required accretion rate shows that this rate is
$\dot{M} \propto (H/R) v_{\rm{turb}} M_{\rm{gas}}^2$. Therefore large gas
masses in a geometrically thick torus can easily provide the accretion
rate needed to trigger and maintain nuclear activity.
The Eq.~(\ref{eq:crittau}) for the transparency also shows that the torus
around an AGN, which runs out of fuel and has less mass in in its
surrounding cloud distribution, will become transparent and eventually
turn into a CND. Activity in a galactic center including a possibly
obscuring torus can also be restarted when additional gas masses
in the form of molecular clouds arrive in the central region.

With the critical gas mass for obscuration ($\tau =1$)
from Eq.~(\ref{eq:crittau}) we can further see that the required
mass accretion rate through the torus is
%-----------------------
\begin{equation}\label{eq:Mdcrit}
  \dot{M} = \frac{\pi}{\sqrt{2}}
  \left(\frac{H}{R}\right)^2 c_s
  \frac{M_{\rm{tot}}}{R}\, .
\end{equation}
%-----------------------
If the inner radius is set by the dust sublimation radius (see below) and
the temperature in clouds is determined by radiative equilibrium with
the central source, the sound speed is constant at the inner radius independent
of the accretion rate or luminosity. With ($R = R_{\rm{in}}
\propto \dot{M}^{1/2}$) we find that the minimum accretion rate for
obscuration is
%-----------------------
\begin{equation}
  \dot{M}_{\rm{obs}} = 3.1 \frac{{\mathrm M}_\odot}{\mathrm yr}\;
  \eta_{-1}^{-1/3} M_7^{2/3}
  \left(\frac{H}{R}\right)^{4/3}
  \left[\frac{c_s}{1.5 \rm{km/s}}\right]^{2/3}
   \, ,
\end{equation}
%-----------------------
where $\eta_{-1}$ is the radiation efficiency at the black hole relative
to 10\% efficiency and $M_7$ the enclosed mass in units $10^7 M_\odot$.
In addition it is assumed that all the mass transported
through the torus actually reaches the black hole.
The required mass accretion rates are highly super-Eddington 
$\dot{M}_{\mathrm{Edd}} = 0.22 M_{\mathrm{BH},7} M_\odot/$yr for the central
black hole with $10\%$ efficiency for transforming rest mass energy into 
radiation. 
The solution is that most of the mass accreted through the torus
is not effectively turn into radiation at the black hole. It is either blown
away in an outflow or jet before reaching the black hole, and thus constitutes
a cooling process for the remaining matter, or the accretion process is indeed
radiatively inefficient at these super-Eddington rates.

Since these collisional disks are stable and long-lived, a possible formation
mechanism of a thick, obscuring torus might be the infall of large gas clumps
onto  a pre-existing circumnuclear ring as investigated
by Vollmer \& Duschl (2002). Infalling gas clouds with different angular
momentum will also lead to increased velocity dispersion in the merged system
as required for a torus.

%__________________________________________________________________________
\subsection{The physics of clouds and the inner boundary of a torus}
In Vollmer \& Duschl (2001a, b) we investigated the physics of gas clouds
in a turbulent or collisional, clumpy disk. It turned out that under realistic
conditions, where the gas is heated by the UV radiation of the central
He{\sc i} star cluster, the gas clouds are stable and self-gravitating.
Their radius is mainly determined by the location of the ionization front
due to the same UV radiation. Under realistic condition this results in
the observed cloud size of $\sim$0.1~pc.
The present study suggests that the gas and dust physics of thick,
obscuring tori is of similar nature.

In this picture two selection effects operate on the initial cloud
mass distribution: (i) only clumps that are dense enough to resist
tidal shear can survive and (ii) clumps that are more massive than
the Jeans mass collapse. In addition, the central UV radiation
field determines the size of the clumps, i.e. their mass for a
given central density. For the CND it turned out that at a
critical distance of $\sim$1.5~pc from the Galactic Center the
clouds, which are dense enough to resist tidal shear, will
collapse. This is in agreement with the observed inner edge of the
CND. If such a mechanism is also at work in a thick, obscuring
torus around an AGN, the inner edge of this torus would be
determined in the same way, if the dust sublimation radius and the
cloud evaporation radius are smaller than the critical radius for
collapse and shear (C\&S-radius) described here.

Dust sublimation due to the radiation of a central source determines
the size of the inner dust free cavity, which will set the source size for NIR
continuum observations (e.g. Wittkowski et al. (1998) for NGC~1068).
This is also the inner radius where cool clouds are shielded by dust opacity.
A possible description of the sublimation radius by Dopita et al. (1998)
%------------------------
\begin{equation}
  R_{\rm sub}=0.84 {\mathrm pc}\,\left[\frac{T_{\rm sub}}{1200 {\rm K}} \right]^{-2.27}
  \,L_{44}^{1/2}\, a_{-5}^{-1/2}\ ,
\end{equation}
%------------------------
shows that $R_{\rm sub}$ is a strong function of
the sublimation temperature $T_{\rm sub}$, which depends on the dust chemistry
and grain sizes with $a_{-5}$ the typical grain radius in
units of $10^{-5}$~cm. The determination of the sublimation radius assumes
radiative equilibrium in the dust sublimation region and is due to
the $R^{-2}$-dilution of the flux from the central luminosity $L_{44}$ in
units of $10^{44}$~erg/s.

If the dust sublimation radius is larger than the C\&S-radius described above,
clouds with masses larger than $\sim$10~M$_{\odot}$ can only survive up to
this radius.
Inside, only the stripped cores of such clouds can survive
(the ``light clouds'' in Vollmer \& Duschl 2001a), because the location
of the ionization front is determined by self-shielding and not dust absorption.

If the cloud evaporation radius is larger than the C\&S-radius, clouds
evaporate before they can reach the C\&S-radius or dust sublimation radius.
Following Krolik \& Begelman (1988) clouds will evaporate when the evaporation time scale equals the accretion time scale at a given distance from the ionizing source, which leads to at a typical evaporation radius of
%------------------------
\begin{equation}\label{R:evap}
  \frac{R_{\rm evap}}{\mathrm{pc}} = \frac{0.79\, M_{\rm{gas},5}\, L_{44}}
  {\dot{M}_{\mathrm{torus}} \, R_{\rm{out}}\, N_{\rm{Cl},24}\, T^{1/2}_5}\ ,
\end{equation}
%------------------------
for a cloud with a gas column density in units of 10~$^{24}$~cm$^{-2}$, 
a temperature at the sonic point of the evaporating flow $T_5$ of $10^{5}$~K, 
a central ionizing luminosity $L_{44}$ in units of $10^{44}$~erg\,s$^{-1}$,
the mass accretion rate $\dot{M}_{\mathrm{torus}}$ measured in M$_\odot$/yr, 
the total gas mass in clouds $M_{\rm{gas},5}$ relative to $10^5$~M$_\odot$, 
and the outer radius of the torus $R_{\rm{out}}$ in parsec. Here we assumed 
$\Sigma \propto R^{-1}$, which is expected for a constant rotation curve 
(the torus residing in the potential of an isothermal star cluster).
Inserting the values for the collisional torus model from 
Table~\ref{tab:resume} one finds $R_{\rm evap} = 0.1$~pc. The evaporation 
radius is sensitive to the ionizing luminosity and $N_{\rm{Cl}}$ 
in an individual cloud,
so that Eq.~(\ref{R:evap}) can be used to set a 
lower limit to the column density at a given radius. Eq.~(\ref{R:evap}) 
seriously underestimates the evaporation radius if it gives
$R_{\rm evap} > R_{\rm{out}}/2$. We can use this
in the collisional model discussed 
here for a central source radiating at the Eddington limit of 
a $3\,10^6$~M$_\odot$ black hole and find that only clouds with
$N_{\rm{Cl}} > 2\,10^{22}$~cm$^{-2}$ will survive when they are exposed 
to the central radiation source. The clouds in the model with 
$r_{\rm{Cl}} = 0.1$~pc and 45~M$_\odot$ are safely above that limit.
At an inner edge of about 2~pc the clouds have a column density of
$\geq 10^{24}$~cm$^{-2}$ (assuming $\phi_V$ a constant) and all clouds 
in the model are stable against evaporation.
%__________________________________________________________________________
%__________________________________________________________________________
\section{Conclusions \label{sec:conclusions}}
We explored three distinct models of clumpy gas disks for the
Circumnuclear Disk (CND) in the Galactic Center and for a
geometrically thick, obscuring torus, as suggested in unified
models of AGN. Three kinds of models were tested: (i) a turbulent,
fully gravitational model, where the energy, which is gained by
mass inflow and differential rotation, is cascaded by turbulence
to smaller scales. Alternatively (ii) the turbulence may be
maintained by the energy input due to supernova explosions, and we
investigated (iii) a collisional model, where the gas viscosity is
due to partly dissipative cloud collisions. In the cases
(i) and (ii) we identify the energy sink to be the
self-gravitation of the gas clouds, which separates them from the
turbulent cascade.

We find that the fully gravitational model cannot be entirely excluded 
for the case of a CND. The cloud masses for the CND model are larger 
than indicated by observations, but the masses in the model depend 
strongly on the assumed outer radius for an equilibrium disk structure.
A turbulent cascade may also exist in a massive torus at
exceedingly large mass accretion rates, where supernovae can drive
the turbulence. Nonetheless the CND and a thick,
obscuring torus are best described by the collisional disk model.
We suggest that both the CND and a thick, obscuring torus share
the same gas physics: they consist of stable, self-gravitating
clouds that are heated by the central UV radiation and the density
of clouds is in both cases close to the critical density with
respect to tidal shear. The mass of single clouds is a few times
10~M$_{\odot}$ in the CND and in a torus.

In this picture a thick, obscuring torus is a more massive CND-like
structure, that is with more gas clumps, and in addition a higher
velocity dispersion.
We show that a torus provides automatically the high mass accretion
rate needed to trigger and maintain nuclear activity and that a geometrically
thick torus is obscuring the view to the center only at super-Eddington rates.

The model for clumpy collisional disks suggests that the appearance as a
CND or an obscuring torus only depends on the mass supply from larger
radii in the galaxy, possibly in the form of pieces of giant molecular clouds.

\acknowledgements{Part of this work was support by the DFG through
grant SFB439 (WJD).}

%---------------------------------------------------------------

\end{document}